\def\rfr#1{eq. (\ref{#1})}

\def\cf#1#2{\dot\Omega^{\rm #2}_{.#1}}

\def\bar{\begin{eqnarray}}
\def\ear{\end{eqnarray}}
\def\bb{\bibitem}
\def\eqi{\begin{equation}}
\def\eqf{\end{equation}}
\def\eqia{\begin{eqnarray}}
\def\eqfa{\end{eqnarray}}
\def\rp#1#2{{#1\over#2}}

\def\lb#1{\label{#1}}

\def\oc2{$\mathcal{O}(c^{-2})$}

\documentclass{aastex}
\usepackage{url}

\RequirePackage{color}

\begin{document}

\title{Advances in the measurement of the Lense-Thirring effect with Satellite Laser Ranging in the gravitational field of the Earth}

\shorttitle{Advances in the measurement of the Lense-Thirring effect with SLR}
\shortauthors{L. Iorio }

\author{Lorenzo Iorio }
\affil{INFN-Sezione di Pisa. Permanent address for correspondence: Viale Unit\`{a} di Italia 68, 70125, Bari (BA), Italy. E-mail: lorenzo.iorio@libero.it}

\begin{abstract}
In this Chapter we deal with the attempts to measure  the  general relativistic gravitomagnetic Lense-Thirring effect with the Satellite Laser Ranging (SLR) technique applied to the existing LAGEOS and LAGEOS II terrestrial satellites and to the recently approved LARES. According to general relativity, the rotation of the central body which acts as source of the gravitational field  changes the spatial orientation of the orbit of test particles determined by the longitude of the ascending node $\Omega$. Such a shift, which is cumulative in time, is of the order of 2 m yr$^{-1}$ in the case of the LAGEOS satellites. Extracting this signature from the data is a demanding task because of many classical orbital perturbations having the same pattern as the gravitomagnetic one, like those induced by the centrifugal oblateness of the Earth which represents a major source of systematic error.
The first issue addressed here is: are the so far published evaluations of the systematic bias induced by  the uncertainty in the even zonal harmonic coefficients $J_{\ell}$ of the multipolar expansion of the Earth's geopotential  reliable and realistic?
The answer is negative. Indeed, if the difference $\Delta J_{\ell}$ among the even zonals estimated in different global solutions (EIGEN-GRACE02S, EIGEN-CG03C, GGM02S, GGM03S, ITG-Grace02s, ITG-Grace03s, JEM01-RL03B, EGM2008) is assumed for the uncertainties $\delta J_{\ell}$  instead of using their more or less calibrated covariance sigmas $\sigma_{J_\ell}$, it turns out that the systematic error $\delta\mu$ in the Lense-Thirring test with the nodes of LAGEOS and LAGEOS II is about 3 to 4 times larger than in the evaluations so far published based on the use of the sigmas of one model at a time separately, amounting up to $37\%$ for the pair EIGEN-GRACE02S/ITG-Grace03s. The comparison among the other recent GRACE-based models yields bias as large as about $25-30\%$.  The major discrepancies still occur for $J_4, J_6$ and $J_8$, which are just the zonals the combined LAGEOS/LAGOES II nodes are most sensitive to.
The second issue is the possibility of reaching a realistic total accuracy of $1\%$ with LAGEOS, LAGEOS II and LARES, which will be launched in the near future.   While LAGEOS and LAGEOS II fly at altitudes of about 6000 km, LARES will be likely placed at an altitude of 1200 km. Thus, it will be sensitive to much more even zonals than LAGEOS and LAGEOS II. Their corrupting impact has been evaluated with the standard Kaula's approach up to degree $\ell=60$ by using $\Delta J_{\ell}$; it turns out that it may be as large as some tens percent. The different orbit of LARES may also have some consequences on the non-gravitational orbital perturbations affecting it which might further degrade the obtainable accuracy.

\end{abstract}

 \section{Introduction}
In the weak-field and slow motion approximation, valid when the magnitude of the gravitational potentials $U$ and velocities $v$ characteristic of the problem under examination are smaller with respect to the speed of light $c$, i.e. for $U/c^2, v/c\lll 1$, the Einstein field equations of general relativity get linearized resembling to the Maxwellian equations of electromagntism. As a consequence, a gravitomagnetic field, induced by the off-diagonal components $g_{0i}, i=1,2,3$ of the space-time metric tensor related to the mass-energy currents of the source of the gravitational field, arises \citep{MashNOVA}; it has no classical counterparts in Newtonian mechanics. The gravitomagnetic field affects orbiting test particles, precessing gyroscopes, moving clocks and atoms and propagating electromagnetic waves \citep{Rug,Scia04}. Perhaps, the most famous gravitomagnetic effects are the precession  of the axis of a gyroscope \citep{Pugh,Schi} and the Lense-Thirring\footnote{According to an interesting historical analysis recently performed in \citep{Pfi07}, it would be more correct to speak about an Einstein-Thirring-Lense effect.} precessions \citep{LT} of the orbit of a test particle, both occurring in the field of a central slowly rotating mass like, e.g., our planet.   Direct, undisputable measurements of such  fundamental predictions of general relativity are not yet available.

The measurement of the gyroscope precession in the Earth's gravitational field has been the goal of the dedicated space-based\footnote{See on the WEB http://einstein.stanford.edu/} GP-B mission \citep{Eve, GPB} launched in 2004 and carrying onboard four superconducting gyroscopes; its data analysis is still ongoing. The target accuracy was originally $1\%$, but it is still unclear if the GP-B team will succeed in reaching such a goal because of some unmodelled effects affecting the gyroscopes: 1) a time variation in the polhode motion of the gyroscopes and 2) very large classical misalignment torques on the gyroscopes.

\begin{figure}
   \includegraphics[width=\columnwidth]{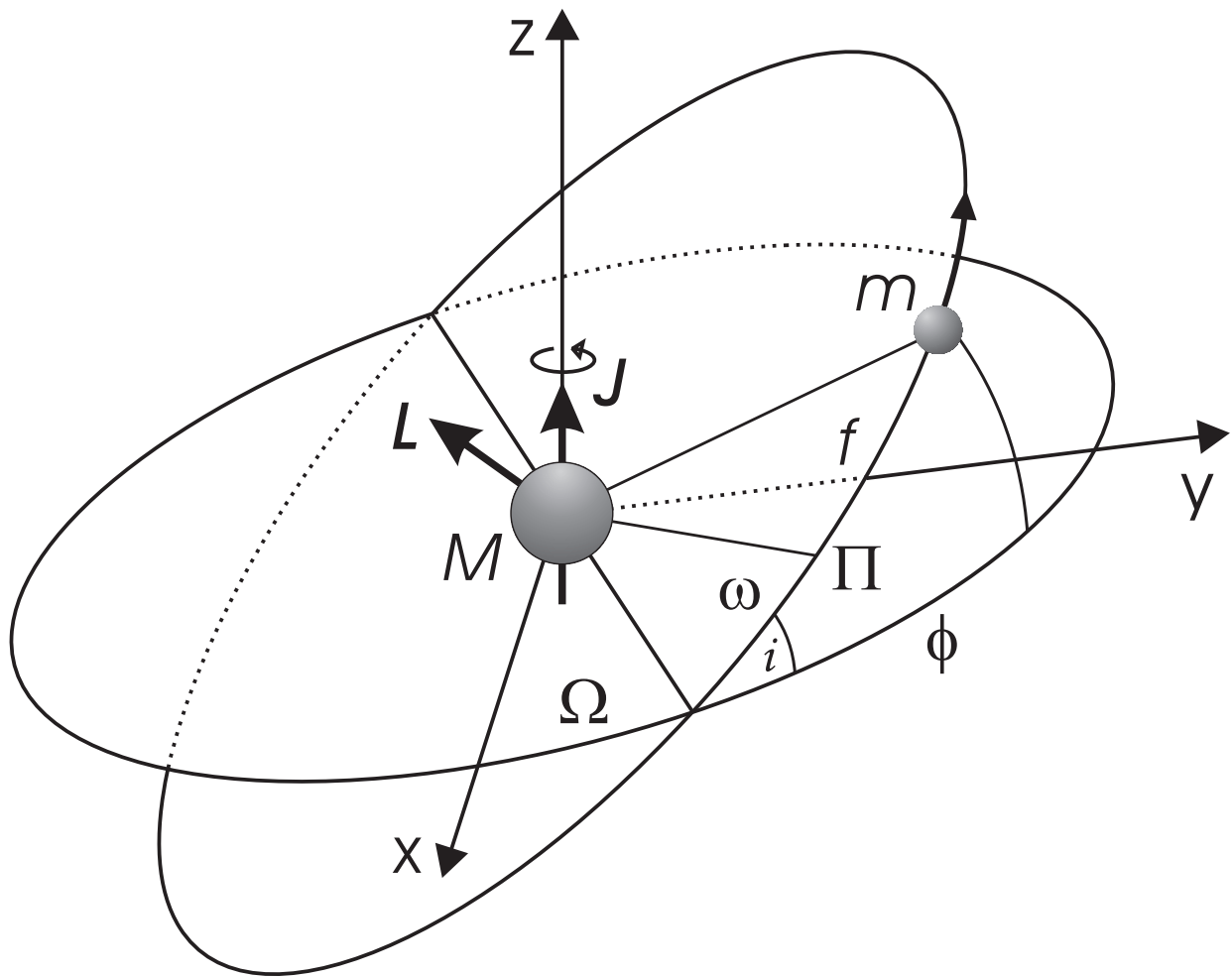}
   \caption{Keplerian orbit. The longitude of the ascending node $\Omega$ is counted from a reference X direction in the reference $\{$XY$\}$ plane, chosen coincident with the equatorial plane of the rotating body of mass $M$ and proper angular momentum $J$, to the line of the nodes, i.e. the intersection of the orbital plane with the reference plane. The argument of pericentre $\omega$ is an angle in the orbital plane counted from the line of the nodes; the location of the pericentre is marked with $\Pi$. The time-dependent position of the moving test particle of mass $m$ is given by the true anomaly $f$, counted anticlockwise from the pericentre; $\phi$ is an azimuthal angle in the $\{$XY$\}$ plane. $L$ is the orbital angular momentum, perpendicular to the orbital plane. The inclination between the orbital plane and the equatorial plane of $M$ is $i$. Courtesy by H. Lichtenegger, OEAW, Graz.}
   \label{plo}
   \end{figure}

In this paper we  will focus on the measurement of the Lense-Thirring effect in the gravitational field of the Earth.
It consists of a secular rate of the longitude of the ascending node $\Omega$ \begin{equation}\dot\Omega_{\rm LT} = \rp{2 G J}{c^2 a^3 (1-e^2)^{3/2}},\lb{let}\end{equation}
and of the argument of pericentre $\omega$
\begin{equation}\dot\omega_{\rm LT} = -\rp{6 G J\cos i}{c^2 a^3 (1-e^2)^{3/2}},\lb{LT_o}\end{equation}
of the orbit of  a test particle. In \rfr{let} and \rfr{LT_o}
$G$ is the Newtonian constant of gravitation, $J$ is the proper angular momentum of the central body, $a$ and $e$ are the semimajor axis and the eccentricity, respectively, of the test particle's orbit and $i$ is its inclination to the central body's equator.  The semimajor axis $a$ determines the size of the ellipse, while its shape is controlled by the eccentricity $0\leq e<1$; an orbit with $e=0$ is a circle. The angles $\Omega$ and $\omega$ fix the orientation of the orbit in the inertial space and in the orbital plane, respectively. $\Omega$, $\omega$ and $i$ can be viewed as the three Euler angles which determine the orientation of a rigid body with respect to an inertial frame. In Figure \ref{plo} we illustrate the geometry of a Keplerian orbit.

In this Chapter we will critically discuss the following two topics
\begin{itemize}
  \item The realistic evaluation in Section \ref{grav} of the total accuracy in the test performed in recent years with the existing Earth's artificial satellites
  LAGEOS and LAGEOS II \citep{Ciu04,Ciu06}. LAGEOS was put into orbit in 1976, followed by its twin LAGEOS II in 1992; they are passive, spherical spacecraft entirely covered by retroreflectors which allow for their accurate tracking through laser pulses sent from Earth-based ground stations according to the Satellite Laser Ranging (SLR) technique. They orbit at altitudes of about 6000 km ($a_{\rm LAGEOS} =12270$ km, $a_{\rm LAGEOS\ II}=12163$ km) in nearly circular paths ($e_{\rm LAGEOS}=0.0045$, $e_{\rm LAGEOS\ II}=0.014$) inclined by 110 deg and 52.65 deg, respectively, to the Earth's equator. The Lense-Thirring effect for their nodes amounts to about 30 milliarcseconds per year (mas yr$^{-1}$) which correspond to about 2 m yr$^{-1}$ at the LAGEOS altitudes.

  The idea of measuring the Lense-Thirring node rate with the just launched LAGEOS satellite, along with the other SLR targets orbiting at that time, was put forth by
\citet{Cug78}. Tests have started to be effectively performed later by using the LAGEOS and LAGEOS II satellites \citep{tanti}, according to a strategy  \citep{Ciu96} involving the use of a suitable linear combination of the nodes $\Omega$ of both satellites and the perigee $\omega$ of LAGEOS II. This was done to reduce the impact of the most relevant source of systematic bias, i.e. the mismodelling in the even ($\ell=2,4,6\ldots$) zonal ($m=0$) harmonic coefficients $J_{\ell}$ of the multipolar expansion of the Newtonian part of the terrestrial gravitational potential due to the diurnal rotation (they induce secular precessions on the node and perigee of a terrestrial satellite much larger than the gravitomagnetic ones. The $J_{\ell}$ coefficients cannot be theoretically computed but must be measured by using dedicated satellites; see Section \ref{grav}): the three-elements combination used allowed for removing the uncertainties in $J_2$ and $J_4$. In \citep{Ciu98} a $\approx 20\%$ test was reported by using the\footnote{Contrary to the subsequent models based on the dedicated satellites CHAMP (http://www-app2.gfz-potsdam.de/pb1/op/champ/index$\_$CHAMP.html) and GRACE (http://www-app2.gfz-potsdam.de/pb1/op/grace/index$\_$GRACE.html), EGM96 relies upon multidecadal tracking of SLR data of a constellation of geodetic satellites including LAGEOS and LAGEOS II as well; thus the possibility of a sort of $a-priori$ `imprinting' of the Lense-Thirring effect itself, not solved-for in EGM96, cannot be neglected.} EGM96 \citep{Lem98} Earth gravity model; subsequent detailed analyses showed that such an evaluation of the total error budget was overly optimistic in view of the likely unreliable computation of the total bias due to the even zonals \citep{Ior03,Ries03a,Ries03b}.  An analogous, huge underestimation turned out to hold also for the effect of the non-gravitational perturbations \citep{Mil87} like the direct solar radiation pressure, the Earth's albedo, various subtle thermal effects depending on  the physical properties of the satellites' surfaces and their rotational state \citep{Inv94,Ves99,Luc01,Luc02,Luc03,Luc04,Lucetal04,Ries03a}, which the perigees  of LAGEOS-like satellites are particularly sensitive to. As a result, the realistic total error budget in the test reported in \citep{Ciu98} might be as large as $60-90\%$ or even more (by considering EGM96 only).

The observable used in \citep{Ciu04} with the GRACE-only EIGEN-GRACE02S model \citep{eigengrace02s} was the following linear combination\footnote{See also \citep{Pav02,Ries03a,Ries03b}.} of the nodes of LAGEOS and LAGEOS II, explicitly computed in \citep{IorMor} following the approach put forth in \citep{Ciu96}
\begin{equation} f=\dot\Omega^{\rm LAGEOS}+
c_1\dot\Omega^{\rm LAGEOS\ II }, \lb{combi}\end{equation}
where \begin{equation} c_1\equiv-\rp{\dot\Omega^{\rm LAGEOS}_{.2}}{\dot\Omega^{\rm
LAGEOS\ II }_{.2}}=-\rp{\cos i_{\rm LAGEOS}}{\cos i_{\rm LAGEOS\
II}}\left(\rp{1-e^2_{\rm LAGEOS\ II}}{1-e^2_{\rm
LAGEOS}}\right)^2\left(\rp{a_{\rm LAGEOS\ II}}{a_{\rm LAGEOS}}\right)^{7/2}.\lb{coff}\end{equation} The coefficients $\dot\Omega_{.\ell}$ of the aliasing classical node precessions \citep{Kau} $\dot\Omega_{\rm class}=\sum_{\ell}\dot\Omega_{.\ell}J_{\ell}$ induced by the even zonals  have been analytically worked out in, e.g. \citep{Ior03}; they yield $c_1=0.544$.
 The Lense-Thirring signature of \rfr{combi} amounts to 47.8 mas yr$^{-1}$. The combination  \rfr{combi} allows, by construction, to remove the aliasing effects due to the static and time-varying parts of the first even zonal $J_2$. The nominal (i.e. computed with the estimated values of $J_{\ell}$, $\ell=4,6...$) bias due to the remaining higher degree even zonals would amount to  about $10^5$ mas yr$^{-1}$; the need of a careful and reliable modeling of such an important source of systematic bias is, thus, quite apparent. Conversely, the nodes of the LAGEOS-type spacecraft are  affected by the non-gravitational accelerations at a $\approx 1\%$ level of the Lense-Thirring effect \citep{Luc01,Luc02,Luc03,Luc04,Lucetal04}. For a comprehensive, up-to-date overview of the numerous and subtle issues concerning the measurement of the Lense-Thirring effect see \citep{IorNOVA}.

  \item  The possibility that the LARES mission, recently approved by the Italian Space Agency (ASI), will be able to measure the Lense-Thirring node precession with an accuracy of the order of $1\%$ (Section \ref{larez}).

      In \citep{vpe76a,vpe76b}  it was proposed to measure the Lense-Thirring precession of the nodes $\Omega$ of a pair of counter-orbiting spacecraft to be launched in terrestrial polar orbits and endowed with drag-free apparatus. A somewhat equivalent, cheaper version of such an idea was put forth in 1986 by \citep{Ciu86} who proposed to launch a passive, geodetic satellite in an orbit identical to that of LAGEOS  apart from the orbital planes which should have been displaced by 180 deg apart. The measurable quantity was, in the case of the proposal by \citet{Ciu86}, the sum of the nodes of LAGEOS and of the new spacecraft, later named LAGEOS III, LARES, WEBER-SAT, in order to cancel to a high level of accuracy the corrupting effect of the multipoles of the Newtonian part of the terrestrial gravitational potential which represent the major source of systematic error (see Section \ref{grav}). Although extensively studied by various groups \citep{CSR,LARES}, such an idea was not implemented for many years.
In \citep{Ioretal02} it was proposed to include also the data from LAGEOS II by using a different observable.  Such an approach was proven in \citep{IorNA} to be potentially useful in making the constraints on the orbital configuration of the new SLR satellite less stringent than it was originally required in view of the recent improvements in our knowledge of the classical part of the terrestrial gravitational potential due to the dedicated CHAMP and, especially, GRACE  missions.

Since reaching high altitudes and minimizing the unavoidable orbital injection errors is expensive, it was explored the possibility of discarding LAGEOS and LAGEOS II using a low-altitude, nearly polar orbit for LARES \citep{LucPao01,Ciu06b}, but in \citep{Ior02,Ior07c} it was proven that such alternative approaches are not feasible. It was also suggested that LARES would be able to probe alternative theories of gravity \citep{Ciu04b}, but also in this case it turned out to be impossible \citep{IorJCAP,Ior07d}.

The stalemate came to an end when ASI recently made the following official announcement  (http://www.asi.it/SiteEN/MotorSearchFullText.aspx?keyw=LARES): ``On February 8, the ASI board approved funding for the LARES mission, that will be launched with VEGA's maiden flight before the end of 2008. LARES is a passive satellite with laser mirrors, and will be used to measure the Lense-Thirring effect.''  The italian version of the announcement yields some more information specifying that LARES, designed in collaboration with National Institute of Nuclear Physics (INFN), is currently under construction by Carlo Gavazzi Space SpA; its Principal Investigator (PI) is I. Ciufolini and its scientific goal is to measure at a $1\%$ level the  Lense-Thirring effect in the gravitational field of the Earth.
Concerning the orbital configuration of LARES, the ASI website says about VEGA that (http://www.asi.it/SiteEN/ContentSite.aspx?Area=Accesso+allo+spazio):
``[...] VEGA can place a 15.000 kg satellite on a low polar orbit, 700 km from the Earth. By lowering the orbit inclination it can launch heavier payloads, whereas diminishing the payload mass it can achieve higher orbits. [...]''    In the latest communication to INFN, Rome, 30 January 2008, \cite{INFN} writes that LARES will be launched with a semimajor axis of approximately 7600 km and an inclination between 60 and 80 deg.
More precise information can be retrieved in Section 5.1, pag 9 of the document Educational Payload on the Vega Maiden Flight
Call For CubeSat Proposals, European Space Agency,
Issue 1
11 February 2008, downloadable at
http://esamultimedia.esa.int/docs/LEX-EC/CubeSat$\%$20CFP$\%$20issue$\%$201.pdf.
    It is  written there that LARES will be launched into a circular orbit with altitude $h=1200$ km, corresponding to a semimajor axis $a_{\rm LARES}=7578$ km, and inclination $i=71$ deg to the Earth's equator.

\end{itemize}

\section{The systematic error of gravitational origin in the LAGEOS-LAGEOS II test}\lb{grav}
The realistic evaluation of the total error budget of the LAGEOS-LAGEOS II node test \citep{Ciu04} raised a lively debate
\citep{Ciu05,Ciu06,IorNA,IorJoG,IorGRG,Ior07,Luc05}, mainly focussed on the impact of the static and time-varying parts of the Newtonian component of the Earth's gravitational potential through the aliasing secular precessions induced on a satellite's node.

In a realistic scenario the path of a probe is not only affected by the gravitomagentic field but also by a huge number of other competing orbital perturbations of gravitational and non-gravitational origin. The most important non-conservative accelerations  \citep{Mil87} are the direct solar radiation pressure, the Earth's albedo and various subtle thermal effects depending on the the physical properties of the satellite's surface and its rotational state \citep{Inv94,Ves99,Luc01,Luc02,Luc03,Luc04,Lucetal04,Ries03a}; however, the nodes  of LAGEOS-like satellites are  sensitive to them at a $\approx 1\%$ level only. Much more important is the impact that the oblateness of the Earth, due to its diurnal rotation, has on the satellite's dynamics.
Indeed, the most insidious perturbations are those induced by the static part of the Newtonian component of the multipolar expansion in spherical harmonics\footnote{The relation among the  even zonals $J_{\ell}$ and the  normalized gravity coefficients $\overline{C}_{\ell 0}$ which are customarily determined in the Earth's gravity models, is $J_{\ell}=-\sqrt{2\ell + 1}\ \overline{C}_{\ell 0}$.} $J_{\ell}, \ell = 2,4,6,...$ of the gravitational potential of the central rotating mass \citep{Kau}: indeed, they affect the node with effects having the same signature of the relativistic signal of interest, i.e. linear trends which are orders of magnitude larger and cannot be removed from the time series of data without affecting the Lense-Thirring pattern itself as well. The only thing that can be done is to model such a corrupting effect as most accurately as possible and assessing, reliably and realistically, the impact of the residual mismodelling on the measurement of the frame-dragging effect.
The secular precessions induced by the even zonals of the geopotential can be written as
\begin{equation}\dot\Omega^{\rm geopot}=\sum_{\ell  =2}\dot\Omega_{.\ell}J_{\ell},\end{equation}
where the coefficients $\dot\Omega_{.\ell}, \ell=2,4,6,...$ depend on the parameters of the Earth ($GM$ and the equatorial radius $R$) and on the semimajor axis $a$, the inclination $i$ and the eccentricity $e$ of the satellite. For example, for $\ell=2$
we have
\begin{equation}\dot\Omega_{.2}=-\rp{3}{2}n\left(\rp{R}{a}\right)^2\rp{\cos i}{(1-e^2)^2};\end{equation}
 $n=\sqrt{GM/a^3}$ is the Keplerian mean motion.
They have been analytically computed up to $\ell=20$  in, e.g., \citep{Ior03}.
Their mismodelling can be written as
\begin{equation}\delta\dot\Omega^{\rm geopot}\leq \sum_{\ell  =2}\left|\dot\Omega_{.\ell}\right|\delta J_{\ell},\lb{mimo}\end{equation}
where $\delta J_{\ell}$ represents our uncertainty in the knowledge of the even zonals $J_{\ell}$

A common feature of all the competing evaluations  so far published is that the systematic bias due to the static component of  the
  geopotential was calculated always by using the released (more or less accurately calibrated) sigmas $\sigma_{J_{\ell}}$  of one Earth gravity model solution at a time for the uncertainties $\delta J_{\ell}$ in the even zonal harmonics, so to say that the model X yields a $x\%$ error, the model Y   yields a $y\%$ error, and so on.

Since a trustable calibration of the formal, statistical uncertainties in the estimated zonals of the covariance matrix of a global solution is always a difficult task to be implemented in a reliable way, a much more realistic and conservative approach consists, instead, of taking the difference\footnote{See Fig.5 of \citep{Luc07} for a comparison of the estimated $\overline{C}_{40}$ in different models.} $\Delta J_{\ell}$ of the estimated even zonals for different pairs of Earth gravity field solutions as representative of the real uncertainty $\delta J_{\ell}$ in the zonals \citep{Lerch}. In Table \ref{tavola1}--Table \ref{tavola8}  we present our results for the most recent GRACE-based models released so far by different institutions and retrievable on the Internet at\footnote{I thank J Ries, CSR, and M Watkins (JPL) for having provided me with the even zonals of the GGM03S \citep{ggm03} and JEM01-RL03B models.}
http://icgem.gfz-potsdam.de/ICGEM/ICGEM.html.   The models used are EIGEN-GRACE02S \citep{eigengrace02s} and EIGEN-CG03C \citep{eigencg03c} from GFZ (Potsdam, Germany), GGM02S \citep{ggm02} and GGM03S \citep{ggm03} from CSR (Austin, Texas), ITG-Grace02s \citep{ITG} and ITG-Grace03s
\citep{itggrace03s} from IGG (Bonn, Germany), JEM01-RL03B from JPL (NASA, USA) and EGM2008 \citep{egm2008} from NGA (USA).
Note that this approach was explicitly followed also by \citet{Ciu96} with the JGM3 and GEMT-2 models.

The systematic bias evaluated with a more realistic approach is about 3 to 4 times larger than one can obtain by only using this or that particular model. The scatter is still quite large and far from the $5-10\%$ claimed in \citep{Ciu04}. In particular, it appears that $J_4$, $J_6$, and to a lesser extent $J_8$, which are just the most relevant zonals for us because of their impact on the combination of \rfr{combi}, are the most uncertain ones, with discrepancies $\Delta J_{\ell}$ between different models, in general, larger than the sum of their sigmas $\sigma_{J_{\ell}}$, calibrated or not.  This is an important feature because the other alternative combinations proposed involving more satellites \citep{IorAji1,Ves} should be less affected since they cancel out the impact of $J_4$ and $J_6$ as well.

Another approach that could be followed to take into account the scatter among the various solutions consists in computing mean and standard deviation of the entire set of values of the even zonals for the models considered so far, degree by degree, and    taking the standard deviations as representative of the uncertainties $\delta J_{\ell}, \ell = 4,6,8,...$. It yields $\delta\mu = 15\%$.

It must be recalled that also the further bias due to the cross-coupling between $J_2$ and the orbit inclination, evaluated to be about $9\%$ in \citep{Ior07}, must be added.
\begin{table}
\caption{\label{tavola1}Impact of the mismodelling in the even zonal harmonics on $f_{\ell}=\left|\dot\Omega^{\rm LAGEOS}_{\ell} + c_1\dot\Omega^{\rm LAGEOS\ II}_{.\ell}\right|\Delta J_{\ell},\ \ell=4,\dots,20$, in mas yr$^{-1}$. Recall that $J_{\ell}=-\sqrt{2\ell + 1}\ \overline{C}_{\ell 0}$; for the uncertainty in the even zonals we have taken here the difference $\Delta\overline{C}_{\ell 0}=\left|\overline{C}_{\ell 0}^{\rm (X)}-\overline{C}_{\ell 0}^{\rm (Y)}\right|$ between the model X $=$ EIGEN-CG03C \citep{eigencg03c} and the model Y $=$ EIGEN-GRACE02S \citep{eigengrace02s}.
EIGEN-CG03C combines data from CHAMP (860 days out of October 2000
to June 2003), GRACE (376 days out of February to
May 2003, July to December 2003 and February to July 2004) and terrestrial measurements; EIGEN-GRACE02S is based on 110 days (out of August and November 2002 and April, May and August 2003) of GRACE-only GPS-GRACE high-low satellite-to-satellite data, on-board measurements of non-gravitational accelerations, and especially GRACE intersatellite tracking data. $\sigma_{\rm X/Y}$ are the covariance calibrated errors for both models. Values of $f_{\ell}$ smaller than 0.1 mas yr$^{-1}$ have not been quoted. The Lense-Thirring precession of the combination of \rfr{combi} amounts to 47.8 mas yr$^{-1}$. The percent bias $\delta\mu$ has been computed by normalizing the linear sum of $f_{\ell}, \ell=4,\dots,20$ to the Lense-Thirring precession. The discrepancies between the models are significant since $\Delta \overline{C}_{\ell 0}$ are larger than the linearly added sigmas for $\ell=4,...16$.}
\begin{tabular}{llll}
\hline
$\ell$ & $\Delta\overline{C}_{\ell 0}$ (EIGEN-CG03C-EIGEN-GRACE02S) & $\sigma_{\rm  X}+\sigma_{\rm Y}$ & $f_{\ell}$  (mas yr$^{-1}$)\\
\hline
4 & $1.96\times 10^{-11}$ &  $1.01\times 10^{-11}$ & 7.3\\
6 & $2.50\times 10^{-11}$ &  $4.8\times 10^{-12}$ & 5.4\\
8 & $4.9\times 10^{-12}$ &  $3.3\times 10^{-12}$ & 0.2\\
10 & $3.7\times 10^{-12}$ &  $3.4\times 10^{-12}$ & -\\
12 & $2.5\times 10^{-12}$ &  $2.3\times 10^{-12}$ & -\\
14 & $6.1\times 10^{-12}$ &  $2.1\times 10^{-12}$ & -\\
16 & $2.1\times 10^{-12}$ &  $1.7\times 10^{-12}$ & -\\
18 & $6\times 10^{-13}$ &  $1.7\times 10^{-12}$ & -\\
20 & $1.7\times 10^{-12}$ &  $1.7\times 10^{-12}$ & -\\
\hline
 &    total bias $\delta\mu = 27\%$ &   \\
\hline
\end{tabular}
 \end{table}
\begin{table}
   \caption{Impact of the mismodelling in the even zonal harmonics as solved for in X=GGM02S \citep{ggm02} and  Y=ITG-Grace02s \citep{ITG}.
   GGM02S is based on 363 days of GRACE-only data   (GPS and intersatellite tracking, neither constraints nor regularization applied)
spread between April 4, 2002 and Dec 31, 2003. The $\sigma$ are formal for both models. $\Delta \overline{C}_{\ell 0}$ are always larger than the linearly added sigmas, apart from   $\ell=12$ and $\ell=18$.}\label{tavola3}
\begin{tabular}{llll}
\hline
$\ell$ & $\Delta\overline{C}_{\ell 0}$ (GGM02S-ITG-Grace02s) & $\sigma_{\rm  X}+\sigma_{\rm Y}$ & $f_{\ell}$  (mas yr$^{-1}$)\\
\hline
4 & $1.9\times 10^{-11}$ &  $8.7\times 10^{-12}$ & 7.2\\
6 & $2.1\times 10^{-11}$ &  $4.6\times 10^{-12}$ & 4.6\\
8 & $5.7\times 10^{-12}$ &  $2.8\times 10^{-12}$ & 0.2\\
10 & $4.5\times 10^{-12}$ &  $2.0\times 10^{-12}$ & -\\
12 & $1.5\times 10^{-12}$ &  $1.8\times 10^{-12}$ & -\\
14 & $6.6\times 10^{-12}$ &  $1.6\times 10^{-12}$ & -\\
16 & $2.9\times 10^{-12}$ &  $1.6\times 10^{-12}$ & -\\
18 & $1.4\times 10^{-12}$ &  $1.6\times 10^{-12}$ & -\\
20 & $2.0\times 10^{-12}$ &  $1.6\times 10^{-12}$ & -\\

\hline
 &   total bias $\delta\mu = 25\%$ &   \\  %
\hline
\end{tabular}
\end{table}
\begin{table}
   \caption{Impact of the mismodelling in the even zonal harmonics as solved for in X=GGM02S \citep{ggm02} and  Y=EIGEN-CG03C \citep{eigencg03c}.
   The $\sigma$ are formal for GGM02S, calibrated for EIGEN-CG03C. $\Delta \overline{C}_{\ell 0}$ are always larger than the linearly added sigmas.}\label{tavola3}
\begin{tabular}{llll}
\hline
$\ell$ & $\Delta\overline{C}_{\ell 0}$ (GGM02S-EIGEN-CG03C) & $\sigma_{\rm  X}+\sigma_{\rm Y}$ & $f_{\ell}$  (mas yr$^{-1}$)\\
\hline
4 & $1.81\times 10^{-11}$ &  $3.7\times 10^{-12}$ & 6.7\\
6 & $1.53\times 10^{-11}$ &  $1.8\times 10^{-12}$ & 3.3\\
8 & $1.5\times 10^{-12}$ &  $1.1\times 10^{-12}$ & -\\
10 & $4.9\times 10^{-12}$ &  $8\times 10^{-13}$ & -\\
12 & $8\times 10^{-13}$ &  $7\times 10^{-13}$ & -\\
14 & $7.7\times 10^{-12}$ &  $6\times 10^{-13}$ & -\\
16 & $3.8\times 10^{-12}$ &  $5\times 10^{-13}$ & -\\
18 & $2.1\times 10^{-12}$ &  $5\times 10^{-13}$ & -\\
20 & $2.3\times 10^{-12}$ &  $4\times 10^{-13}$ & -\\

\hline
 &   total bias $\delta\mu = 22\%$ &   \\  %
\hline
\end{tabular}
\end{table}

\begin{table}
   \caption{Bias due to the mismodelling in the even zonals of the models X=ITG-Grace03s \citep{itggrace03s}, based on GRACE-only accumulated normal equations from data out of September 2002-April 2007 (neither apriori information nor regularization used), and Y=GGM02S \citep{ggm02}.  The $\sigma$ for both models are formal. $\Delta \overline{C}_{\ell 0}$ are always larger than the linearly added sigmas, apart from  $\ell=12$ and $\ell=18$.}\label{tavola11}
\begin{tabular}{llll}
\hline
$\ell$ & $\Delta\overline{C}_{\ell 0}$ (ITG-Grace03s-GGM02S) & $\sigma_{\rm  X}+\sigma_{\rm Y}$ & $f_{\ell}$  (mas yr$^{-1}$)\\
\hline
4 & $2.58\times 10^{-11}$ &  $8.6\times 10^{-12}$ & 9.6\\
6 & $1.39\times 10^{-11}$ &  $4.7\times 10^{-12}$ & 3.1\\
8 & $5.6\times 10^{-12}$ &  $2.9\times 10^{-12}$ & 0.2\\
10 & $1.03\times 10^{-11}$ &  $2\times 10^{-12}$ & -\\
12 & $7\times 10^{-13}$ &  $1.8\times 10^{-12}$ & -\\
14 & $7.3\times 10^{-12}$ &  $1.6\times 10^{-12}$ & -\\
16 & $2.6\times 10^{-12}$ &  $1.6\times 10^{-12}$ & -\\
18 & $8\times 10^{-13}$ &  $1.6\times 10^{-12}$ & -\\
20 & $2.4\times 10^{-12}$ &  $1.6\times 10^{-12}$ & -\\

\hline
&   total bias $\delta\mu = 27\%$ &   \\  %
\hline
\end{tabular}
\end{table}
\begin{table}
   \caption{Bias due to the mismodelling in the even zonals of the models  X = GGM02S \citep{ggm02} and Y = GGM03S \citep{ggm03} retrieved from data spanning January 2003 to December 2006.
    The $\sigma$ for GGM03S are calibrated. $\Delta \overline{C}_{\ell 0}$ are larger than the linearly added sigmas for $\ell = 4,6$. (The other zonals are of no concern)}\label{tavola03S}
\begin{tabular}{llll}
\hline
$\ell$ & $\Delta\overline{C}_{\ell 0}$ (GGM02S-GGM03S) & $\sigma_{\rm  X}+\sigma_{\rm Y}$ & $f_{\ell}$  (mas yr$^{-1}$)\\
\hline
4 & $1.87\times 10^{-11}$ &  $1.25\times 10^{-11}$ & 6.9\\
6 & $1.96\times 10^{-11}$ &  $6.7\times 10^{-12}$ & 4.2\\
8 & $3.8\times 10^{-12}$ &  $4.3\times 10^{-12}$ & 0.1\\
10 & $8.9\times 10^{-12}$ &  $2.8\times 10^{-12}$ & 0.1\\
12 & $6\times 10^{-13}$ &  $2.4\times 10^{-12}$ & -\\
14 & $6.6\times 10^{-12}$ &  $2.1\times 10^{-12}$ & -\\
16 & $2.1\times 10^{-12}$ &  $2.0\times 10^{-12}$ & -\\
18 & $1.8\times 10^{-12}$ &  $2.0\times 10^{-12}$ & -\\
20 & $2.2\times 10^{-12}$ &  $1.9\times 10^{-12}$ & -\\

\hline
&   total bias $\delta\mu = 24\%$ &   \\  %
\hline
\end{tabular}
\end{table}
\begin{table}
   \caption{Bias due to the mismodelling in the even zonals of the models  X = EIGEN-GRACE02S \citep{eigengrace02s} and Y = GGM03S \citep{ggm03}.
    The $\sigma$ for both models are calibrated. $\Delta \overline{C}_{\ell 0}$ are always larger than the linearly added sigmas apart from $\ell = 14,18$.}\label{tavola033S}
 \begin{tabular}{llll}
\hline
$\ell$ & $\Delta\overline{C}_{\ell 0}$ (EIGEN-GRACE02S-GGM03S) & $\sigma_{\rm  X}+\sigma_{\rm Y}$ & $f_{\ell}$  (mas yr$^{-1}$)\\
\hline
4 & $2.00\times 10^{-11}$ &  $8.1\times 10^{-12}$ & 7.4\\
6 & $2.92\times 10^{-11}$ &  $4.3\times 10^{-12}$ & 6.3\\
8 & $1.05\times 10^{-11}$ &  $3.0\times 10^{-12}$ & 0.4\\
10 & $7.8\times 10^{-12}$ &  $2.9\times 10^{-12}$ & 0.1\\
12 & $3.9\times 10^{-12}$ &  $1.8\times 10^{-12}$ & -\\
14 & $5\times 10^{-13}$ &  $1.7\times 10^{-12}$ & -\\
16 & $1.7\times 10^{-12}$ &  $1.4\times 10^{-12}$ & -\\
18 & $2\times 10^{-13}$ &  $1.4\times 10^{-12}$ & -\\
20 & $2.5\times 10^{-12}$ &  $1.4\times 10^{-12}$ & -\\

\hline
&   total bias $\delta\mu = 30\%$ &   \\  %
\hline
\end{tabular}
\end{table}
\begin{table}
   \caption{Bias due to the mismodelling in the even zonals of the models  X = JEM01-RL03B, based on 49 months of GRACE-only data, and Y = GGM03S \citep{ggm03}.
    The $\sigma$ for GGM03S are calibrated. $\Delta \overline{C}_{\ell 0}$ are always larger than the linearly added sigmas apart from $\ell = 16$.}\label{tavolaJEM1}
\begin{tabular}{llll}
\hline
$\ell$ & $\Delta\overline{C}_{\ell 0}$ (JEM01-RL03B-GGM03S) & $\sigma_{\rm  X}+\sigma_{\rm Y}$ & $f_{\ell}$  (mas yr$^{-1}$)\\
\hline
4 & $1.97\times 10^{-11}$ &  $4.3\times 10^{-12}$ & 7.3\\
6 & $2.7\times 10^{-12}$ &  $2.3\times 10^{-12}$ & 0.6\\
8 & $1.7\times 10^{-12}$ &  $1.6\times 10^{-12}$ & -\\
10 & $2.3\times 10^{-12}$ &  $8\times 10^{-13}$ & -\\
12 & $7\times 10^{-13}$ &  $7\times 10^{-13}$ & -\\
14 & $1.0\times 10^{-12}$ &  $6\times 10^{-13}$ & -\\
16 & $2\times 10^{-13}$ &  $5\times 10^{-13}$ & -\\
18 & $7\times 10^{-13}$ &  $5\times 10^{-13}$ & -\\
20 & $5\times 10^{-13}$ &  $4\times 10^{-13}$ & -\\

\hline
&   total bias $\delta\mu = 17\%$ &   \\  %
\hline
\end{tabular}
\end{table}
\begin{table}
   \caption{Bias due to the mismodelling in the even zonals of the models  X = JEM01-RL03B and Y = ITG-Grace03s \citep{itggrace03s}.
    The $\sigma$ for ITG-Grace03s are formal. $\Delta \overline{C}_{\ell 0}$ are always larger than the linearly added sigmas.}\label{tavolaJEM2}
\begin{tabular}{llll}
\hline
$\ell$ & $\Delta\overline{C}_{\ell 0}$ (JEM01-RL03B-ITG-Grace03s) & $\sigma_{\rm  X}+\sigma_{\rm Y}$ & $f_{\ell}$  (mas yr$^{-1}$)\\
\hline
4 & $2.68\times 10^{-11}$ &  $4\times 10^{-13}$ & 9.9\\
6 & $3.0\times 10^{-12}$ &  $2\times 10^{-13}$ & 0.6\\
8 & $3.4\times 10^{-12}$ &  $1\times 10^{-13}$ & 0.1\\
10 & $3.6\times 10^{-12}$ &  $1\times 10^{-13}$ & -\\
12 & $6\times 10^{-13}$ &  $9\times 10^{-14}$ & -\\
14 & $1.7\times 10^{-12}$ &  $9\times 10^{-14}$ & -\\
16 & $4\times 10^{-13}$ &  $8\times 10^{-14}$ & -\\
18 & $4\times 10^{-13}$ &  $8\times 10^{-14}$ & -\\
20 & $7\times 10^{-13}$ &  $8\times 10^{-14}$ & -\\

\hline
&   total bias $\delta\mu = 22\%$ &   \\  %
\hline
\end{tabular}
\end{table}
\begin{table}
   \caption{Aliasing effect of the mismodelling in the even zonal harmonics estimated in the X=ITG-Grace03s \citep{itggrace03s} and the Y=EIGEN-GRACE02S \citep{eigengrace02s} models.  The covariance matrix $\sigma$ for ITG-Grace03s are formal, while the ones of EIGEN-GRACE02S are calibrated. $\Delta \overline{C}_{\ell 0}$ are larger than the linearly added sigmas for $\ell =4,...,20$, apart from $\ell=18$. }\label{tavola7}
\begin{tabular}{llll}
\hline
$\ell$ & $\Delta\overline{C}_{\ell 0}$ (ITG-Grace03s-EIGEN-GRACE02S) & $\sigma_{\rm  X}+\sigma_{\rm Y}$ & $f_{\ell}$  (mas yr$^{-1}$)\\
\hline
4 & $2.72\times 10^{-11}$ &  $3.9\times 10^{-12}$ & 10.1\\
6 & $2.35\times 10^{-11}$ &  $2.0\times 10^{-12}$ & 5.1\\
8 & $1.23\times 10^{-11}$ &  $1.5\times 10^{-12}$ & 0.4\\
10 & $9.2\times 10^{-12}$ &  $2.1\times 10^{-12}$ & 0.1\\
12 & $4.1\times 10^{-12}$ &  $1.2\times 10^{-12}$ & -\\
14 & $5.8\times 10^{-12}$ &  $1.2\times 10^{-12}$ & -\\
16 & $3.4\times 10^{-12}$ &  $9\times 10^{-13}$ & -\\
18 & $5\times 10^{-13}$ &  $1.0\times 10^{-12}$ & -\\
20 & $1.8\times 10^{-12}$ &  $1.1\times 10^{-12}$ & -\\

\hline
 &   total bias $\delta\mu = 37\%$ &   \\  %
\hline
\end{tabular}

\end{table}
\begin{table}
   \caption{Impact of the mismodelling in the even zonal harmonics estimated in the X=EGM2008 \citep{egm2008} and the Y=EIGEN-GRACE02S \citep{eigengrace02s} models.  The covariance matrix $\sigma$ are calibrated for both EGM2008 and EIGEN-GRACE02S. $\Delta \overline{C}_{\ell 0}$ are larger than the linearly added sigmas for $\ell =4,...,20$, apart from $\ell=18$. }\label{tavola8}
\begin{tabular}{llll}
\hline
$\ell$ & $\Delta\overline{C}_{\ell 0}$ (EGM2008-EIGEN-GRACE02S) & $\sigma_{\rm  X}+\sigma_{\rm Y}$ & $f_{\ell}$  (mas yr$^{-1}$)\\
\hline
4 & $2.71\times 10^{-11}$ &  $8.3\times 10^{-12}$ & 10.0\\
6 & $2.35\times 10^{-11}$ &  $4.1\times 10^{-12}$ & 5.0\\
8 & $1.23\times 10^{-11}$ &  $2.7\times 10^{-12}$ & 0.4\\
10 & $9.2\times 10^{-12}$ &  $2.9\times 10^{-12}$ & 0.1\\
12 & $4.1\times 10^{-12}$ &  $1.9\times 10^{-12}$ & -\\
14 & $5.8\times 10^{-12}$ &  $1.8\times 10^{-12}$ & -\\
16 & $3.4\times 10^{-12}$ &  $1.5\times 10^{-12}$ & -\\
18 & $5\times 10^{-13}$ &  $1.5\times 10^{-12}$ & -\\
20 & $1.8\times 10^{-12}$ &  $1.5\times 10^{-12}$ & -\\

\hline
 &   total bias $\delta\mu = 33\%$ &   \\  %
\hline
\end{tabular}

\end{table}

\section{A conservative evaluation of the impact of the geopotential on the LARES mission}\lb{larez}
The combination which will be used for measuring the Lense-Thirring effect with LAGEOS, LAGEOS II and LARES is  \citep{IorNA}
\eqi \dot\Omega^{\rm LAGEOS}+k_1\dot\Omega^{\rm LAGEOS\ II}+ k_2\dot\Omega^{\rm LARES}.
\lb{combaz}\eqf
The coefficients $k_1$ and $k_2$ entering \rfr{combaz}      are defined as
\begin{equation}
\begin{array}{lll}
k_1 = \rp{\cf 2{LARES}\cf4{LAGEOS}-\cf 2{LAGEOS}\cf 4{LARES}}{\cf 2{LAGEOS\ II}\cf 4{LARES}-\cf 2{LARES}\cf 4{LAGEOS\ II}}=0.3697,\\\\
k_2 =  \rp{\cf 2{LAGEOS}\cf4{LAGEOS\ II}-\cf 2{LAGEOS\ II}\cf 4{LAGEOS}}{\cf 2{LAGEOS\ II}\cf 4{LARES}-\cf 2{LARES}\cf 4{LAGEOS\ II}}=0.0619.
\end{array}\lb{cofis}
 \end{equation}
The combination \rfr{combaz} cancels out, by construction, the impact of the first two even zonals.
The total Lense-Thirring effect, according to \rfr{combaz} and \rfr{cofis}, amounts to 47.8 mas yr$^{-1}$.
The systematic error due to the uncancelled even zonals $J_6, J_8,...$ can be conservatively evaluated as
\eqi\delta\mu\leq \sum_{\ell = 6}\left|\dot\Omega^{\rm LAGEOS}_{.\ell}+k_1\dot\Omega^{\rm LAGEOS\ II}_{.\ell}+ k_2\dot\Omega^{\rm LARES}_{.\ell}\right|\delta J_{\ell}\lb{biass}\eqf

Of crucial importance is how to assess $\delta J_{\ell}$.  By proceeding as in Section \ref{grav} and by using the same models up to degree $\ell = 60$  because of the lower altitude of LARES with respect to LAGEOS and LAGEOS II which brings into play more even zonals, we have the results presented in Table \ref{tavolay}.
 \begin{table}\caption{ Systematic percent error $\delta\mu$ in the measurement of the Lense-Thirring effect with LAGEOS, LAGEOS II and LARES according to \rfr{biass} and $\delta J_{\ell}= \Delta J_{\ell}$ up to degree $\ell = 60$ for the global Earth's gravity solutions  considered here.}\label{tavolay}

\footnotesize{\begin{tabular}{ll}
\hline
Models compared & $\delta\mu(\%)$\\
\hline
EGM2008$-$JEM01-RL03B & 8\%\\
EGM2008$-$GGM02S & 27\%\\
EGM2008$-$GGM03S & 5\%\\
EGM2008$-$ITG-Grace02 & 4\%\\
EGM2008$-$ITG-Grace03 & 0.1\%\\
EGM2008$-$EIGEN-CG03C & 38\%\\
EGM2008$-$EIGEN-GRACE02S & 53\%\\
%
%
JEM01-RL03B$-$GGM02S & $28\%$\\
JEM01-RL03B$-$GGM03S & $10\%$\\
JEM01-RL03B$-$ITG-Grace02 & $8\%$\\
JEM01-RL03B$-$ITG-Grace03s & $8\%$\\
JEM01-RL03B$-$EIGEN-CG03C & $44\%$\\
JEM01-RL03B$-$EIGEN-GRACE02S & $57\%$\\
%
%
GGM02S$-$GGM03S & $24\%$\\
GGM02S$-$ITG-Grace02& $28\%$\\
GGM02S$-$ITG-Grace03s& $26\%$\\
GGM02S$-$EIGEN-CG03C & $27\%$\\
GGM02S$-$EIGEN-GRACE02S & $36\%$\\
%
%
GGM03S$-$ITG-Grace02 & $5\%$\\
GGM03S$-$ITG-Grace03s & $5\%$\\
GGM03S$-$EIGEN-CG03C & $36\%$\\
GGM03S$-$EIGEN-GRACE02S & $52\%$\\
%
%
ITG-Grace02$-$ITG-Grace03s & $4\%$\\
ITG-Grace02$-$EIGEN-CG03C & $39\%$\\
ITG-Grace02$-$EIGEN-GRACE02S & $54\%$\\
%
%
ITG-Grace03s$-$EIGEN-CG03C & $38\%$\\
ITG-Grace03s$-$EIGEN-GRACE02S & $53\%$\\
%
%
EIGEN-CG03C$-$EIGEN-GRACE02S & $27\%$\\
%
%
%
\hline

\end{tabular}
}
\end{table}

It must be stressed that they may be still optimistic: indeed, computations for $\ell > 60$ become unreliable because of numerical instability of the results (obtained with two different softwares).

If, instead, one assumes $\delta J_{\ell}=s_{\ell},\ \ell=2,4,6,...$ i.e., the standard deviations of the sets of all the best estimates of $J_{\ell}$ for the models considered here the systematic bias, up to $\ell=60$, amounts to $26\%$. Again, also this result may turn out to be optimistic for the same reasons as before.


It must be pointed out that the evaluations presented here rely upon calculations of the coefficients $\dot\Omega_{.\ell}$ performed with the well known standard approach by Kaula \cite{Kau}; it would be important to try to follow also different computational strategies in order to test them.

It is worthwhile noting that also the impact of the subtle non-gravitational perturbations will be different with respect to the original proposal because LARES will fly in a different and lower orbit and its thermal behavior will  probably be different with respect to  LAGEOS and LAGEOS II.  The reduction of the impact of the thermal accelerations, like the Yarkowsky-Schach effects, should have  been reached with two concentric spheres. However, as explained in \cite{Andres}, this solution will increase the floating potential of LARES because of the much higher electrical resistivity and, thus, the perturbative effects produced by the charged particle drag. Moreover, drag will increase also because of the lower orbit of the satellite, both in its neutral and charged components. Also the Earth's albedo, with its anisotropic components, should have a major effect.

Another point which must be considered is the realistic orbit accuracy obtainable for LARES. Indeed, at a lower orbit the normal points RMS will be probably higher with respect to the present RMS obtained for the two LAGEOS satellites (a few mm), as we presently know for the Stella and Starlette normal points. Of course, such an accuracy is a function of several aspects.

\section{Conclusions}
In this Chapter we have shown how the so far published evaluations of the total systematic error in the Lense-Thirring measurement with the combined nodes of the SLR LAGEOS and LAGEOS II satellites due to the classical node precessions induced by the even zonal harmonics of the geopotential  are  optimistic. Indeed, they are all based on the use of the covariance matrix's sigmas, more or less reliably calibrated, of various Earth gravity model solutions used one at a time separately in such a way that the model X yields an error of $x\%$, the model Y yields an error $y\%$, etc. Instead,  comparing the estimated values of the even zonals for different pairs of models   allows for a much more realistic evaluation of the real uncertainties in our knowledge of the static part of the geopotential. As a consequence, the bias in the Lense-Thirring effect measurement is about $3-4$ times larger than that so far claimed, amounting to various tens percent ($37\%$ for the pair EIGEN-GRACE02S and ITG-GRACE03s, about $25-30\%$ for the other most recent GRACE-based solutions).

Applying the same strategy to the ongoing LARES mission shows that the goal of reaching a $1\%$ measurement of the Lense-Thirring effect with LAGEOS, LAGEOS II and LARES is optimistic. Indeed, since LARES will orbit at a lower altitude with respect to the LAGEOS satellites, a larger number of even zonal harmonics are to be taken into account. Assessing realistically their impact is not easy. Straightforward calculations up to degree $\ell = 60$ with the standard Kaula's approach yield errors as large as some tens percent. Such an important point certainly deserves great attention. Another issue which may potentially degrade the expected accuracy is the impact of some non-gravitational perturbations which would have a larger effect on LARES than expected because of its lower orbit.


 \end{document}